\documentclass[12pt]{article}
\usepackage{amsfonts}

\newcommand{\Nl}{\mathbb{N}}
\newcommand{\Ir}{\mathbb{Z}}
\newcommand{\Cx}{\mathbb{C}}
\newcommand{\A}{\mathcal{A}}
\newcommand{\HH}{\mathcal{H}}
\newcommand{\Section}[1]%
{\section{#1}\setcounter{equation}{0}%
\setcounter{theorem}{0}}
\newtheorem{theorem}{Theorem}
\newtheorem{lemma}[theorem]{Lemma}

\newcommand{\eq}[1]{(\ref{#1})}
\newenvironment{proof}{\noindent {\bf Proof: }}{\QED\medskip}
\def\QED{{\hspace*{\fill}{\vrule height 1.8ex width 1.8ex }\quad} 
    \vskip 0pt plus20pt}
\newcommand{\ret}{\nonumber \\}
\newcommand{\Aloc}{\A_{\mbox{\tiny loc}}}
\newcommand{\pprime}{{\prime\prime}}
\def\idty{{\mathchoice {\rm 1\mskip-4mu l} {\rm 1\mskip-4mu l} %
{\rm 1\mskip-4.5mu l} {\rm 1\mskip-5mu l}}}
\newcommand{\be}{\begin{equation}}
\newcommand{\ee}{\end{equation}}
\newcommand{\bea}{\begin{eqnarray}}
\newcommand{\eea}{\end{eqnarray}}
\newcommand{\beann}{\begin{eqnarray*}}
\newcommand{\eeann}{\end{eqnarray*}}

\newcommand{\ket}[1]{\left\vert #1\right\rangle}
\newcommand{\bra}[1]{\left\langle #1\right\vert}
\newcommand{\wslim}{{\rm w}^*\mbox{-}\lim}
\begin{document}
\thispagestyle{empty}
\noindent
Archived as {\tt cond-mat/9709208} and {\tt mp\_arc 97-511}
\hspace{\fill}\\
September 1997; minor revisions April 13, 1998\hspace{\fill}
\vspace{7pt}
\begin{center}
{\LARGE\bf The complete set of ground states} \\[8pt] 
{\LARGE\bf of the ferromagnetic XXZ chains} \\[10pt]
Tohru Koma\\
Department of Physics\\
Gakushuin University\\
Mejiro, Toshima-ku\\
Tokyo 171, JAPAN\\
E-mail: {\tt koma@riron.gakushuin.ac.jp}\\[8pt]
Bruno Nachtergaele\\
Department of Mathematics\\ 
University of California, Davis\\
Davis, CA 95611-8633, USA\\
E-mail: {\tt bxn@math.ucdavis.edu}\\[10pt]

{\bf Abstract}\\[3pt]  
\end{center}  

We show that the well-known translation invariant ground states and the
recently discovered kink and antikink ground states are the complete set of
pure infinite-volume ground states (in the sense of local stability) of the 
spin-$S$ ferromagnetic XXZ chains  with Hamiltonian 
$H=-\sum_x[S^{(1)}_xS^{(1)}_{x+1}+S^{(2)}_xS^{(2)}_{x+1} +\Delta
S^{(3)}_xS^{(3)}_{x+1}]$, for all $\Delta >1$, and all $S\in\frac{1}{2}\Nl$.
For the isotropic model ($\Delta =1$) we show that all ground states are
translation invariant.

For the proof of these statements we propose a strategy for demonstrating
completeness of the list of the pure  infinite-volume ground states of a
quantum many-body system, of which the present results for the XXX and XXZ
chains  can be seen as an example. The result for $\Delta>1$ can also be proved
by  an easy extension to general $S$  of the method used in [T.~Matsui, Lett.
Math. Phys. {\bf 37}  (1996) 397] for the spin-1/2 ferromagnetic XXZ chain with
$\Delta>1$. However, our proof is different and does not rely on the existence
of a spectral gap. In particular, it also  works to prove absence of 
non-translationally invariant ground states for the isotropic chains
($\Delta=1$), which have a gapless excitation spectrum. 

Our results show that, while any small amount of  the anisotropy is enough to
stabilize the domain walls against the quantum  fluctuations, no boundary
condition exists that would  stabilize a domain wall in the isotropic  model
$(\Delta=1)$. 

\bigskip

\noindent {\bf Keywords:} quantum Heisenberg chain, Heisenberg-Ising
chain, XXZ chain, ferromagnets, kink ground states, domain walls, 
pure ground states, completeness, local stability.

\vspace{\fill}
\hrule width2truein
\smallskip
{\baselineskip=12pt
\noindent
\copyright\ 1997, 1998 T. Koma and B. Nachtergaele. Reproduction of
this article, by any means, is permitted for non-commercial purposes.
\par}
\newpage
\Section{Introduction}

In this paper we consider the ground states of 
the spin-$S$ ferromagnetic XXZ chains with formal Hamiltonian 
\be
H=-\sum_x\left[S^{(1)}_xS^{(1)}_{x+1}+S^{(2)}_xS^{(2)}_{x+1}
+\Delta S^{(3)}_xS^{(3)}_{x+1}\right]
\label{model}
\ee
for all $\Delta\geq 1$, and all $S\in\frac{1}{2}\Nl$.  Our main objective is
to show that the known ground states of these models are the {\it complete}
set of ground states in the sense of stability under local perturbations. In
the isotropic case ($\Delta=1$) all ground states are translation invariant. 
For the anisotropic ferromagnetic XXZ chains ($\Delta >1$), non-translation
invariant ground states (kinks and antikinks) were recently discovered,
independently by Alcaraz, Salinas, and Wreszinski \cite{ASW} and Gottstein 
and Werner \cite{GW}. See Section \ref{sec:aux_results} for a detailed 
description of these states. 

Before we describe in detail our results, we explain the somewhat surprising
insights that the detailed  study of the XXZ ferromagnets has brought along.
We believe that from this  and other simple models one can obtain interesting
information about the  nature and the effects of quantum fluctuations,
with relevance for a  variety of phenomena in magnetism and interface
physics.

In the absence of an external field and at low temperatures real 
ferromagnets usually exhibit domain wall structures. But it is   commonly
believed that pinning effects due to impurities and defects  are crucial to
stabilize domain walls against quantum fluctuations
\cite{Schilling,PSega,MMR,LGon,HHC}. In other words, in models of quantum
ferromagnetism  that do not incorporate impurities or defects, one should not
expect to find stable domain walls because they are destroyed by quantum
fluctuations. While pinning may still be important to keep the domain walls
from wandering around, the results for the XXZ Heisenberg model mentioned
above, show that an arbitrary small amount of anisotropy is sometimes enough
to stabilize interfaces against the quantum  fluctuations.

A natural question, which is the subject of this paper, is whether or not one
can obtain other ground states, e.g., that have interfaces with a different
geometry or internal structure. Is it possible to construct more infinite-volume
ground states, e.g., by imposing radically different boundary  conditions
and taking thermodynamic limits?
The standard mathematical formulation of this question is in terms of the local
stability condition: what is the complete set of infinite-volume ground states 
$\omega$ satisfying
\be
\lim_{\Lambda\to\Ir}\omega(A^* [H_\Lambda,A])
\geq 0,\quad \mbox{for all }A\in \Aloc\; ?
\label{gs_intro}\ee
The notations are explained and more details are given in Section 
\ref{sec:main_results}. See \cite{FW} for an informative discussion of the 
problem of obtaining all states by imposing different boundary conditions. 

The first result in this direction is due to Matsui \cite{Mat}. For
the spin-1/2 XXZ chain with $\Delta >1$, he proved that any pure 
infinite-volume ground state is either the all spins up, the all spins down, 
a kink, or an antikink state, i.e., the set of states given in \cite{ASW} 
and \cite{GW} is the complete set. Matsui's proof, however, relies on rather
special properties  of the model. In particular, his method cannot be used to
treat the isotropic (XXX) ferromagnetic chain which has a gapless  excitation
spectrum. The most important contribution in this paper is the proof that all 
the ground states of the XXX ferromagnetic chain are  translation invariant. 
Our strategy also gives a new proof of completeness in the case of the XXZ
chains, for all values of the spin. We also believe that our method can be
adapted to other models.  Absence of non-translation invariant ground
states for the AKLT model \cite{AKLT} was recently proved in \cite{Mat2}.

It is often said that the local stability definition of ground state
\eq{gs_intro},  although natural both from the theoretical and experimental 
point of view, is  of little practical use, because, until recently, it seemed
impossible to find all its solution for a non-trivial quantum model.  Even in
cases where there is general agreement about what the full set of ground states
should be, it was impossible to show that the known set of ground states is
indeed  the complete set (see the comments at the end of Section 6.2 in volume
2  of \cite{BraRo}). It remains true that extracting interesting information
from the local stability inequalities alone, is usually quite difficult. But we
hope to convince the reader that by combining it with other ideas (zero-energy
states) and by using equivalent formulations, such as the one due to Bratteli,
Kishimoto, and Robinson (see Theorem \ref{thm:BKR}), interesting progress can
be made.

Our result for the XXX chains should not be
confused with another, very interesting,  completeness result for the XXX
ferromagnetic chain. In \cite{BT} Babbitt and Thomas proved that the Bethe
Ansatz states lead to a complete  resolution of the identity (Plancherel
formula) in the GNS representation  of any of its pure translation invariant
ground states. These two completeness results nicely complement each other.

The present paper is organized as follows:  In Section~\ref{sec:main_results}
we present our main results  (Theorem \ref{thm:mainXXZ} for the anisotropic
and Theorem \ref{thm:mainXXX} for the isotropic chains).  We also give a
schematic description of the proof. In the proofs we use some properties of
the finite-volume and infinite-volume  zero-energy ground states of the XXZ
chains.  These are presented in Section \ref{sec:aux_results}. Section 
\ref{sec:proofXXZ} is devoted to the proof of Theorem \ref{thm:mainXXZ}.
The proof of Theorem \ref{thm:mainXXX} is in Section \ref{sec:proofXXX}.

\Section{Main results and structure of the proof}
\label{sec:main_results}

In this section we limit ourselves to present a precise statement of the
main results and an outline of the proof. More detailed definitions and
notations, as well as additional results, are discussed in the next
section. 

A natural way of defining a quantum spin model is by specifying
its Heisenberg dynamics, or, equivalently, the generator of this
dynamics, $\delta$, given as a densely defined closed operator on
the algebra of quasi-local observables of the spin chain. For a
chain of spin-$S$ degrees of freedom, the algebra of quasi-local
observables $\A$ is the C*-inductive limit ($\equiv$ the completion 
for the operator norm) of the strictly local observables 
\be
\Aloc=\bigcup_{\Lambda\subset\Ir}\A_\Lambda,\quad
\A_\Lambda=\bigotimes_{x\in\Lambda}M_\Cx(2S+1),
\ee
where the union is over finite subsets of the integers, and $M_\Cx(2S+1)$
denotes the complex $(2S+1)\times(2S+1)$ matrices.

The local Hamiltonians and the basis observables are usually expressed in
terms of the spin matrices $S^{(1)}_x, S^{(2)}_x, S^{(3)}_x$, which are
the generators of the $(2S+1)$-dimensional irreducible unitary 
representation of SU(2). The subscript $x\in \Ir$ refers to the site in
the chain, i.e., the factor in the tensor product, with which these
matrices are associated. The local Hamiltonians for the ferromagnetic XXZ
chains are given by
\be
-\sum_{x=-L}^{L-1}\left[S_x^{(1)}S_{x+1}^{(1)}+S_x^{(2)}S_{x+1}^{(2)}
+\Delta S_x^{(3)}S_{x+1}^{(3)}\right]=H_L\in\A_{[-L,L]}. 
\ee
Here $\Delta$ is the anisotropy parameter and we will always assume
$\Delta\geq 1$. The generator of the dynamics is then defined by
\begin{equation}
\delta(A)=\lim_{L\to\infty}[H_L, A]
\end{equation}
in the sense that
\begin{equation}
A(t)=\exp(it\delta)A,\quad \mbox{for all }A\in\A. 
\end{equation}

By definition, a state $\omega$, i.e., a positive normalized linear 
functional, on $\A$ is a {\it ground state} of the model iff
\be
\omega(A^*\delta(A))=\lim_{L\to\infty}\omega(A^* [H_L,A])
\geq 0,\quad \mbox{for all }A\in \A.
\label{gs}\ee
This inequality expresses the property that any local perturbation of
$\omega$ has a total energy at least equal to the energy of the
unperturbed state. So, ground states minimize the energy locally.

The solutions of \eq{gs} can be described as follows:

\begin{enumerate}
\item 
If $\Delta =1$ (the isotropic case), all solutions are
translation invariant. The support of the finite-volume restrictions of
any ground state is contained in the subspace of maximal total spin. This
means that these states are, in fact, permutation invariant. In the case
$S=1/2$, they are all permutation invariant states. For $S>1/2$, there
are permutation invariant states that are not ground states.
\item 
If $\Delta >1$, there are two translation invariant ground 
states, $\omega_\uparrow$ and $\omega_\downarrow$, which 
are characterized by the property
\be
\omega_\uparrow(S_x^{(3)})=+S,\quad\omega_\downarrow(S_x^{(3)})=-S,
\mbox{ for all } x\in\Ir. 
\label{up_and_down}\ee
In addition to these two, there are two infinite sets of pure
non-translation invariant ground states, called kinks and antikinks.
The kink ground states approach $\omega_\uparrow$ as $x\to-\infty$ and
$\omega_\downarrow$ as $x\to+\infty$. For antikink states the roles of plus
and minus infinity are reversed.
\end{enumerate}

These ground states were all known from the previous works 
\cite{ASW,GW}. For more detailed properties we refer to the next section. 
In this paper we prove that they are the complete set of solutions of \eq{gs}. 
This is the content of the following two theorems:

\begin{theorem} \label{thm:mainXXZ}
For the spin-S XXZ ferromagnetic chain with the anisotropic  coupling
$\Delta>1$, the following statements are valid:  There are two translation
invariant pure ground states, namely $\omega_\uparrow$ and $\omega_\downarrow$
determined by \eq{up_and_down}. Any pure infinite-volume ground state that is
not translation invariant is either a kink, or an antikink ground state,
belonging to the set described in \cite{GW}.
\end{theorem}

\begin{theorem} \label{thm:mainXXX}
In the spin-S ferromagnetic Heisenberg chain with the isotropic 
coupling $\Delta=1$, all ground states (i.e., all solutions of \eq{gs})
are translation invariant.
\end{theorem}

By combining these two theorems with previously known results,  in particular
\cite{GW}, one can also prove that any pure infinite-volume ground state can be
obtained as a thermodynamic limit of finite-volume states determined by
eigenvectors of the finite-volume Hamiltonians with simple boundary terms. 
E.g., for $\Delta >1$, any pure infinite-volume ground state $\omega$ that
is not translation invariant satisfies
\be
\wslim_{n\to\infty}\omega\circ\tau_{-n}=\omega_\uparrow
\mbox{ and \ } \wslim_{n\to\infty}\omega\circ\tau_n=\omega_\downarrow,
\label{1stomega}
\ee
or
\be
\wslim_{n\to\infty}\omega\circ\tau_{-n}=\omega_\downarrow
\mbox{ and \ } \wslim_{n\to\infty}\omega\circ\tau_n=\omega_\uparrow,
\label{2ndomega}
\ee
where $\omega_\uparrow$ and $\omega_\downarrow$ are the states defined in
\eq{up_and_down}. Here $\tau_j$ is the lattice translation by $j$ sites,
i.e., $\tau_j(A)\in \A_{\Lambda+j}$, for all $A\in\A_\Lambda$.
For any state $\omega$, $\omega\circ\tau_j$ is the state obtained 
by translation of $\omega$ over $j$ lattice spacings.
In the first case (\ref{1stomega}), $\omega$ is the thermodynamic limit of
finite-volume ground states  of the local Hamiltonians 
${\tilde H}_\Lambda^{+-}$ defined in \eq{Ham_pm}. In the second case 
(\ref{2ndomega}), the same holds with the local Hamiltonians 
${\tilde H}_\Lambda^{-+}$. For a more detailed version of 
Theorem~\ref{thm:mainXXZ} see  Section~\ref{sec:proofXXZ}. 

Next, we give a brief outline of the proof of Theorem \ref{thm:mainXXZ}.
The steps are formulated in a more general setting than just the
XXZ chain, because this makes clear how one might generalize 
the result. In general, each step is a non-trivial problem, but in the
case of the XXZ chain we have completed each of them.
For a description of the proof of Theorem \ref{thm:mainXXX}, which
is closely related to the proof of Theorem \ref{thm:mainXXZ}, see Section
\ref{sec:proofXXX}.
\medskip

{\it Step 1.} The first step is to determine the possible asymptotic behavior.
This is relatively easy for one-dimensional systems. One only needs
to determine the following sets of w$^*$-limit points of 
arbitrary ground states:
\bea
LP_+&:=& \mbox{w$^*$-limit points of } \{\omega\circ\tau_n\mid \omega 
\mbox{ satisfies } \eq{gs}, \mbox{ and } n\geq 0\},\\
LP_-&:=& \mbox{w$^*$-limit points of } \{\omega\circ\tau_n\mid \omega 
\mbox{ satisfies } \eq{gs}, \mbox{ and } n\leq 0\}.
\eea
In reflection symmetric models such as the XXZ chain, $LP_-=LP_+=:LP$,
and we can, for simplicity, assume we are in this situation. In general,
$LP$ is a convex set of states. For convenience, let us also assume 
it is a finite-dimensional simplex consisting of translation
invariant ground states of the model. In the case of the XXZ model
with $\Delta>1$ this is the case and $LP$ is the segment 
$\{t\omega_\uparrow + (1-t)\omega_\downarrow\mid 0\leq t\leq 1\}$. 
For the isotropic case $LP$ is also the set of translation invariant
ground states, but the set of extreme points (the pure states) is infinite 
due to the continuous rotation
symmetry. Everything below can be adapted to handle also this case.
Denote by $LP_0$ the set of pure states in $LP$.
\medskip

{\it Step 2.} Next, construct orthogonal projections $P_\xi\in\Aloc$,
$\xi\in LP_0$, such that 
\begin{equation}
\xi(P_\eta)\approx \delta_{\xi,\eta},\quad \hbox{ for all }\xi,\eta\in LP_0 .
\label{orthogonalP}
\end{equation} 
Such projections always exist on the basis of general abstract arguments
\cite{BraRo}, and one can assume that $\sum_\xi P_\xi=\idty$.
\medskip

{\it Step 3.} Then one shows that for all pairs of pure 
states $\xi,\eta\in LP_0$, the finite volume
Hamiltonian can be written in the form
\begin{equation}
\tilde{H}_{[-L,L]}^{(\xi,\eta)}=\sum_{x=-L-1}^L h^{(\xi,\eta)}_{x,x+1}
+ W^{(\xi,\eta)}_L
\end{equation}
for a suitable choice of the  interaction $h^{(\xi,\eta)}_{x,x+1}\geq 0$ and
boundary terms $W^{(\xi,\eta)}_L$, such that there is a ground 
state $\omega^{(\xi,\eta)}$ of the model with left and right asymptotics
given by $\xi$ and $\eta$ respectively, and such that 
$\omega^{(\xi,\eta)}(h^{(\xi,\eta)}_{x,x+1})=0$, for all $x\in \Ir$,
i.e., $\omega^{(\xi,\eta)}$ is a {\it zero-energy ground state}
of the set of local Hamiltonians ${\tilde H}_{[-L,L]}^{(\xi,\eta)}$ in the
sense of \cite{GW}. We assume that the complete set of zero-energy
ground states of $\{{\tilde H}_{[-L,L]}^{(\xi,\eta)}\}$ is known, which is
the case for the XXX and XXZ chains.
For the XXZ chain the $\omega^{(\xi,\eta)}$ are the known ground states
described above.
The aim is to prove that any ground state $\omega$ in the sense
of \eq{gs} is a convex combination of zero-energy ground states.
\medskip

{\it Step 4.}. An arbitrary ground state can now be decomposed according
to its left and right asymptotics, using the projections satisfying 
(\ref{orthogonalP}):
\begin{equation}
\omega(\, \cdots\,)\approx \sum_{\xi,\eta \in LP_0} 
\omega(\tau_{-a}(P_\xi)(\,\cdots\,)\tau_{b}(P_\eta)),\quad a,b\gg 1. 
\end{equation}
Next, we use the following theorem due to Bratteli, Kishimoto, and
Robinson \cite{BKR}.

\begin{theorem}\label{thm:BKR}
Let $\tilde{H}_\Lambda$ be a set of local
Hamiltonians of the form $\tilde{H}_\Lambda=H_{\Lambda\cup\partial\Lambda}
+W_\Lambda$ 
where $W_\Lambda \in \A_{\Lambda^c}$ are arbitrary boundary terms, 
and $\partial\Lambda\subset\Lambda^c$.
A state $\omega$ is a ground state in the sense of \eq{gs}
iff $\omega$ satisfies 
\be
\omega(\tilde{H}_\Lambda)=\inf\left\{\omega'(\tilde{H}_\Lambda)\mid
\omega'\in C_\Lambda^\omega\right\}
\label{1.5}
\ee
for all finite $\Lambda$, and where 
\begin{equation}
C_\Lambda^\omega:=\left\{\omega^\prime, \mbox{a state on } \A\mid
\omega^\prime(A)=\omega(A) \mbox{ for all } A \in \A_{\Lambda^c}\right\}. 
\end{equation}
\end{theorem}

The difficult part is to construct suitable $\omega^\prime\in 
C_\Lambda^\omega$, and to prove an energy estimate for them.
This is done by inserting a piece of the state $\omega^{(\xi,\eta)}$ into 
$\omega(\tau_{-a}(P_\xi)(\,\cdots\,)\tau_{b}(P_\eta))$ appearing
in the decomposition according to left and right asymptotics,
and by using the appropriate boundary terms.
By applying Theorem \ref{thm:BKR}, one then obtains that the states
$\omega(\tau_{-a}(P_\xi)(\,\cdots\,)\tau_{b}(P_\eta))$ satisfy
\begin{equation}
\lim_{a,b\to\infty}\omega(\tau_{-a}(P_\xi) h^{(\xi,\eta)}_{x,x+1}
\tau_{b}(P_\eta)) =0
\end{equation}
for all $x\in \Ir$. In combination with Step 3, it then follows that 
$\omega$ is a convex combination of zero-energy ground states. 

{From} the outline above it is clear that the notion zero-energy ground
states plays an important role in our completeness proof. 
We expect, however, that weaker notions can be substituted for it.
E.g., the zero-energy condition is only used asymptotically,
i.e., in the thermodynamic limit. A similar approach was reported 
by Matsui \cite{Mat3}.

\Section{Notations and Auxiliary Results}
\label{sec:aux_results}

\subsection{Notations}

We consider the standard ferromagnetic XXZ chain with arbitrary spin. 
The volumes $\Lambda$ we consider will always be finite unions of 
intervals $[a,b]\subset\Ir$, and the half-infinite intervals of the 
form $[a,+\infty)$ and $(-\infty,b]$. The boundary of
$\Lambda$, $\partial\Lambda$, is defined by
\begin{equation}
\partial\Lambda=\{x\in\Ir\setminus\Lambda\mid \{x-1,x+1\}\cap\Lambda
\neq\emptyset\}. 
\end{equation}
E.g., if $\Lambda=[a,b]$, $\partial\Lambda=\{a-1,b+1\}$,
and $\Lambda\cup\partial\Lambda=[a-1,b+1]$.
The finite volume Hamiltonians of the XXZ Heisenberg chain are
given by
\be
H_\Lambda=\sum_{x=a}^{b-1} h_{x,x+1}
\label{hamS}
\ee
for all $a<b\in\Ir$, and where
\be
h_{x,x+1}:= -\Delta^{-1}\left[
S_x^{(1)}S_{x+1}^{(1)}+S_x^{(2)}S_{x+1}^{(2)}\right]
-\left[S_x^{(3)}S_{x+1}^{(3)}-S^2\right],\mbox{ for all }x\in\Ir. 
\label{localHam}
\ee
Here $\Delta\geq 1$ is the anisotropic coupling, and
$S=1/2,1,3/2,\ldots$. The normalization of the interactions \eq{localHam} 
is such that one can consider the limit $\Delta\rightarrow\infty$ 
without difficulty. In this limit the models become the ferromagnetic 
classical Ising chains with spin-$S$. If $\Lambda_1\cap\Lambda_2=\emptyset$, 
$H_{\Lambda_1\cup\Lambda_2}=H_{\Lambda_1}+H_{\Lambda_2}$.
This determines the Hamiltonians for all $\Lambda$ that are finite
unions of finite intervals. 

For the description of the kink and the antikink ground states
we introduce two other Hamiltonians with specific boundary terms 
as follows: for any $\Lambda=[a,b]$, 
\begin{equation}
\tilde{H}_\Lambda^{+-}=H_{\Lambda\cup\partial\Lambda} 
+ A(\Delta)\left[S_{b+1}^{(3)}-S_{a-1}^{(3)}\right],
\label{Ham_pm}
\end{equation}
and
\begin{equation}
\tilde{H}_\Lambda^{-+}=H_{\Lambda\cup\partial\Lambda} 
- A(\Delta)\left[S_{b+1}^{(3)}-S_{a-1}^{(3)}\right],
\label{Ham_mp}
\end{equation}
where $A(\Delta):=S\sqrt{1-\Delta^{-2}}$. 
We will use the convention that tildes indicate that the local
Hamiltonians include boundary terms. Typically, one has that
$\tilde{H}_\Lambda\in\A_{\Lambda\cup\partial\Lambda}$, while
$H_\Lambda\in\A_\Lambda$. The particular boundary terms in \eq{Ham_pm}
and \eq{Ham_mp} make the quantum group $SU_q(2)$ symmetry that
the XXZ chains possess explicit \cite{PS}. These are the local Hamiltonians
studied in \cite{ASW,GW,KN1}. For convenience we also put 
\be
\tilde{H}_\Lambda^{++}=\tilde{H}_\Lambda^{--}=H_{\Lambda\cup\partial\Lambda}
\label{Ham_ppmm}
\ee
We will also use the notation
\bea
h^{+-}_{x,x+1}&=&-\Delta^{-1}\left[
S_x^{(1)}S_{x+1}^{(1)}+S_x^{(2)}S_{x+1}^{(2)}\right]
-\left[S_x^{(3)}S_{x+1}^{(3)}-S^2\right]
+ A(\Delta)\left[S_{x+1}^{(3)}-S_x^{(3)}\right]\label{h+-},\\
h^{-+}_{x,x+1}&=&-\Delta^{-1}\left[
S_x^{(1)}S_{x+1}^{(1)}+S_x^{(2)}S_{x+1}^{(2)}\right]
-\left[S_x^{(3)}S_{x+1}^{(3)}-S^2\right]
- A(\Delta)\left[S_{x+1}^{(3)}-S_x^{(3)}\right]\label{h-+},
\eea
and
\be
h^{++}_{x,x+1}=h^{--}_{x,x+1}=h_{x,x+1}
\label{h++--}.
\ee
Note that for $\alpha,\beta=\pm$, we have
\be
\tilde{H}_{[a,b]}^{\alpha\beta}=\sum_{x=a-1}^b h^{\alpha\beta}_{x,x+1}. 
\label{localHam_pm}
\ee
In the case of $\alpha\beta=+-$ and $\alpha\beta=-+$, the boundary terms in 
corresponding Hamiltonians allow for ground states with a kink, 
or an antikink \cite{ASW,GW}. 

\subsection{Properties of the ground states}

For the proof of Theorem \ref{thm:mainXXZ} we will use 
detailed properties of the known ground states of the XXZ chain.
In this section we bring together the properties known in the
literature and prove some additional results that we need.

\subsubsection{Finite-volume ground states}
\label{fintevolumeGS}

Consider the spin-S XXZ ferromagnetic chain given by 
\begin{equation}
\tilde{H}_\Lambda^{\alpha\beta}= \sum_{x=a}^{b-1} h_{x,x+1}^{\alpha\beta}
\label{hamL}
\end{equation}
with the anisotropic coupling $\Delta >1$ and with 
$\Lambda\cup\partial\Lambda=[a,b]$, and where $h_{x,x+1}^{\alpha\beta}$
has been defined in (\ref{h+-}--\ref{h++--}). For concreteness we consider
the case $\alpha=-,\beta=+$. The case $\alpha=+, \beta=-$ is identical 
up to a reflection. The cases $\alpha=\beta=+$, and $\alpha=\beta=-$
are trivial. The ground state given by Alcaraz, Salinas and Wreszinski 
\cite{ASW} is 
\begin{equation}
\Phi_{[a,b]}^{(M)}=\sum_{m_x}\prod_{x\in[a,b]}q^{x(S-m_x)}
w(m_x)|\{m_x\}\rangle, 
\label{ASWGS}
\end{equation}
where $m_x$ is the third component of the spin at the site $x$ 
with the range $-S\le m_x\le S$, and the sum is restricted to 
the configurations $\{m_x\}$ of the spins for which 
$\sum_x m_x=M$. The parameter $q$ is defined by $1<\Delta=(q+q^{-1})/2$ 
with $0<q<1$, and the weights $w(m_x)$ are given by 
\begin{equation}
w(m)=\sqrt{\frac{(2S)!}{(S-m)!(S+m)!}}. 
\end{equation}
One can easily show that $h_{x,x+1}^{-+}\Phi_{[a,b]}^{(M)}=0$ 
for any $x\in [a,b-1]$. Thus we have 
${\tilde H}_\Lambda^{-+}\Phi_{[a,b]}^{(M)}=0$. Since $h_{x,x+1}^{-+}\ge 0$, 
the vector $\Phi_{[a,b]}^{(M)}$ is a ground state 
of the Hamiltonian $\tilde{H}_\Lambda^{-+}$ of (\ref{hamL}). 
Further this ground state $\Phi_{[a,b]}^{(M)}$ is unique in the sector 
of the fixed magnetization $M$ from the Perron-Frobenius theorem. 

Following \cite{GW}, we introduce a vector 
\begin{eqnarray}
{\check \Psi}_{[a,b]}^{-+}:=\sum_M
\Phi_{[a,b]}^{(M)}=\bigotimes_{x\in[a,b]}{\check \chi}_x
\end{eqnarray}
with 
\begin{equation}
{\check \chi}_x:=\sum_{m_x=-S}^Sq^{x(S-m_x)}w(m_x)|m_x\rangle. 
\end{equation}
The norm of ${\check \chi}_x$ can be easily computed as 
\begin{equation}
\Vert{\check \chi}_x\Vert=\left(1+q^{2x}\right)^S. 
\end{equation}
Therefore the normalized vector is 
\begin{eqnarray}
\Psi_{[a,b]}^{-+}:=\bigotimes_{x\in[a,b]}\chi_x
\label{defkinkvector1}
\end{eqnarray}
with 
\begin{equation}
\chi_x:=\frac{1}{\left(1+q^{2x}\right)^S}
\sum_{m_x=-S}^Sq^{x(S-m_x)}w(m_x)|m_x\rangle. 
\label{defkinkvector2}
\end{equation}
We denote by $\psi^{-+}_{[a,b]}$ the corresponding state, i.e., 
the expectation 
\begin{equation}
\psi_{[a,b]}^{-+}(\cdots)=
\left\langle\Psi_{[a,b]}^{-+},(\cdots)\Psi_{[a,b]}^{-+}\right\rangle. 
\label{defkinkpsi}
\end{equation}
Clearly this is an antikink with the center at the origin. Similarly we 
can construct the kink $\psi_{[a,b]}^{+-}$ which is the reflection 
of the antikink centered at the origin. 

The probability of finding $S_x^{(3)}\ne\alpha S$ at the site $x$ is 
given by 
\begin{equation}
{\rm Prob}_{[a,b]}^{-+}\left(S_x^{(3)}\ne \alpha S\right):=
\psi_{[a,b]}^{-+}(\idty-P_x^\alpha), \quad \mbox{for} \ \alpha=\pm, 
\end{equation}
where $P_x^\pm$ are the projections onto the state with the spin 
$S_x^{(3)}=S$ at the site $x$ for the plus sign, and the state with 
$S_x^{(3)}=-S$ for the minus sign. Similarly we define 
by ${\rm Prob}_{[a,b]}^{+-}\left(S_x^{(3)}\ne \alpha S\right)$ 
the corresponding probabilities for the kink $\psi_{[a,b]}^{+-}$. 

\begin{lemma}\label{lem:probnotS}
The following estimates are valid: 
\begin{equation}
{\rm Prob}_{[a,b]}^{-+}\left(S_x^{(3)}\ne\alpha S\right)\le 2Sq^{2\alpha x},
\quad \mbox{for} \ \alpha=\pm, 
\end{equation}
\begin{equation}
{\rm Prob}_{[a,b]}^{+-}\left(S_x^{(3)}\ne\alpha S\right)\le 2Sq^{-2\alpha x},
\quad \mbox{for} \ \alpha=\pm. 
\end{equation}
\end{lemma}

\begin{proof}
The probability of finding $S_x^{(3)}=S$ at the site $x$ is given by 
\begin{equation}
{\rm Prob}_{[a,b]}^{-+}\left(S_x^{(3)}=S\right):=
\psi_{[a,b]}^{-+}(P_x^+)=\frac{1}{(1+q^{2x})^{2S}}  
\end{equation}
from (\ref{defkinkvector1}), (\ref{defkinkvector2}), and 
(\ref{defkinkpsi}). Therefore 
\begin{equation}
{\rm Prob}_{[a,b]}^{-+}\left(S_x^{(3)}\ne S\right)=
1-\frac{1}{(1+q^{2x})^{2S}}\le 1-e^{-2Sq^{2x}}\le 2Sq^{2x},
\end{equation}
where we have used the following two inequalities: 
$1+u\le e^u$ for $u\ge 0$, and $1-e^{-v}\le v$ for $v\ge 0$. 
The rest are treated in the same way. 
\end{proof}

\subsubsection{Infinite-volume ground states}
\label{sec:infinite_volume}

All pure ground states --at this point, we should rather say, all
{\it known} pure ground states--, of the ferromagnetic XXZ chains 
happen to be  zero-energy ground states, i.e., if $\omega$ is
a pure ground state of a spin-$S$ ferromagnetic XXZ chain, then
there is a choice of $\alpha,\beta=\pm$ such that
\be
\omega(h^{\alpha\beta}_{x,x+1})=0,\quad\mbox{ for all }x\in\Ir.
\label{zero_energy}
\ee
It is known from the work of Gottstein and Werner \cite{GW} 
(and some straightforward generalizations thereof)
that the following list describes {\it all} 
zero-energy ground states of these models:

\begin{itemize}
\item If $\Delta=1$, all zero-energy ground states are translation
invariant and the pure zero-energy ground states are all rotations
of the state determined by
\begin{equation}
\omega_\uparrow(S^{(3)}_x)=S,\quad\mbox{ for all }x\in\Ir.
\end{equation}
This means that there is $g\in $ SU(2) such that 
\begin{equation}
\omega(A)=\omega_\uparrow(R_g(A)),\quad\mbox{ for all }A\in\A,
\end{equation}
where the rotation automorphism $R_g$ is determined by 
\begin{equation}
R_g(A)=\bigotimes_{x\in\Lambda} U_x^*(g)AU_x(g), 
\quad\mbox{ for all }A\in\A_{\Lambda},
\end{equation}
where $U_x^*(g)$ is the $(2S+1)$-dimensional irreducible unitary
representation of SU(2) acting on the state space at site $x$.
It is then clear that any zero-energy ground state of the isotropic
chain is invariant under arbitrary permutations of the sites of
the chain.
\item If $\Delta>1$, as mentioned before, not all zero-energy
ground states are translation invariant.
\begin{enumerate}
\item There are two translation invariant zero-energy ground states,
$\omega_\uparrow$ and $\omega_\downarrow$, respectively determined
by 
\begin{equation}
\omega_\uparrow(S^{(3)}_x)=S,\mbox{ and } \omega_\downarrow(S^{(3)}_x)=-S,
\quad\mbox{ for all }x\in\Ir.
\end{equation}
\item There is an infinite family of pure {\it kink} ground states,
which are all obtained as thermodynamic limits of the finite-volume
ground states of the Hamiltonians ${\tilde H}^{+-}_\Lambda$,
described in the previous paragraphs.
For a more detailed description of the resulting space of states
see \cite{GW}. The kink ground states satisfy
\begin{equation}
\omega(h^{+-}_{x,x+1})=0 .
\end{equation}
Note that the translation invariant ground states have zero energy
for each of the interactions $h^{\alpha\beta}_{x,x+1}$.
\item There is an infinite family of pure {\it antikink} ground 
states which can be obtained from the kink ground states either 
by reflection (interchanging left and right), or by a rotation
over $\pi$ about an axis in the XY plane (spin flip, interchanging up
and down). All antikink ground states satisfy
\begin{equation}
\omega(h^{-+}_{x,x+1})=0 .
\end{equation}
\end{enumerate}
\end{itemize}

Gottstein and Werner \cite{GW} also prove that if $\omega$ is
a pure zero-energy ground state of the XXZ chains with local Hamiltonians 
$\tilde{H}_\Lambda^{-+}$, then exactly one of the following
must be true:
\begin{itemize}
\item Either $\omega$ is translation invariant, and it is then
either $\omega_\uparrow$ or $\omega_\downarrow$,
\item or $\omega$ is  represented by a unit vector in the GNS Hilbert 
space of the state $\omega_0$ obtained by the following thermodynamic 
limit:
\be
\omega_0(A)=\lim_{b\to +\infty}\lim_{a\to -\infty}\psi^{-+}_{[a,b]}(A)
\ee
which describes an antikink centered at the origin. 
Here $\psi_{[a,b]}^{-+}$ is given by (\ref{defkinkpsi}). 
For a description of the GNS representation $\pi_{\omega_0}$
of $\omega_0$, and the GNS Hilbert space $\HH_{\omega_0}$, see \cite{GW} or 
\cite{KN1}. So, in this case there is a vector $\psi\in\HH_{\omega_0}$,
such that for all $A\in\A$
\be
\omega(A)=\langle \psi,\pi_{\omega_0}(A)\psi\rangle \; .
\ee
\end{itemize}
Moreover, $\psi$ belongs to the kernel of the GNS Hamiltonian.
Together with Theorem \ref{thm:mainXXZbis}, this result (and its analogue for kink states)
also proves that all infinite-volume ground states  of the XXZ chains are
thermodynamic limits of finite-volume ground states of local Hamiltonians with
one-site boundary terms as, e.g.,  the $\tilde{H}_\Lambda^{\alpha\beta}$
defined in (\ref{Ham_mp}--\ref{Ham_ppmm}).

\section[PTXXZ]{Proof of Theorem \ref{thm:mainXXZ}}
\label{sec:proofXXZ}
\setcounter{equation}{0}%
\setcounter{theorem}{0}%

We begin with three short lemmas that will be used in the proof of
Theorem \ref{thm:mainXXZ} and Theorem \ref{thm:mainXXX}.

\begin{lemma}\label{lem:finite_energy}
Let ${\tilde H}_{[a,b]}=\sum_{x=a-1}^bh_{x,x+1}$ be local
Hamiltonians with a translation invariant interaction satisfying 
$0\leq h_{x,x+1}\leq h\idty$, and suppose there exists a state $\eta$
of $\A$ such that $\eta(h_{x,x+1})=0$, for all $x\in\Ir$. Then,
if $\omega$ is a ground state, i.e., $\omega$ satisfies \eq{gs},
we have
\be
0\leq\omega({\tilde H}_{[a,b]})\leq 2h
\label{energy_bound}\ee
for all $a<b\in \Ir$, and
\be
\lim_{x\to\pm\infty}\omega(h_{x,x+1})=0.
\label{limit0}
\ee
\end{lemma}

\begin{proof}
Since $h_{x,x+1}\ge 0$, we get the lower bound. 
To obtain the upper bound, we consider the trial state 
\be
\omega'=\omega_{[a,b]^c}\otimes \eta_{[a,b]},
\ee
where the subscripts denote the restriction to the algebra 
on the corresponding subvolume. Then, by Theorem \ref{thm:BKR}, we have 
\be
\omega\left({\tilde H}_{[a,b]}\right)\le\omega'\left({\tilde H}_{[a,b]}\right)=
\left(\omega_{[a,b]^c}\otimes \eta_{[a,b]}\right)
(h_{a-1,a}+h_{b,b+1})\le 2h. 
\ee
This proves \eq{energy_bound}. {From} \eq{energy_bound} it follows 
that $\lim\omega({\tilde H}_{[a,b]})$ is a convergent sum of non-negative 
terms terms and, hence, \eq{limit0} is obvious.
\end{proof}

\begin{lemma}\label{lem:asymtotic_states}
Suppose $a_n\in\Ir$ such that either $\lim_n a_n =+\infty$ or
$\lim_n a_n=-\infty$, and suppose $\omega$ is a ground state
of the spin$-S$ XXZ chain with $\Delta >1$. If there exists 
\be
\lim_n \omega\circ\tau_{a_n}=\eta,
\label{limitn}
\ee
then there exists $t\in [0,1]$ such that
\begin{equation}
\eta=t\omega_\uparrow + (1-t)\omega_\downarrow,
\end{equation}
where $\omega_\uparrow$ and $\omega_\downarrow$ are the ``up'' and ``down'' 
ground states defined in \eq{up_and_down}.
\end{lemma}

\begin{proof}
By Lemma \ref{lem:finite_energy} we have that $\eta$ is a 
zero-energy ground state with 
\begin{equation}
\eta(h_{x,x+1})=0,\mbox{ for all } x\in\Ir.
\end{equation}
This implies that $\eta$ is a convex combination of the two translation 
invariant zero-energy ground states $\omega_\uparrow$ and $\omega_\downarrow$.
\end{proof}

We will use the following elementary lemma in our estimates.

\begin{lemma}\label{lem:state_bound}
Let $\omega$ be a state of a C*-algebra $\A$, and suppose $P\in\A$ is
an orthogonal projection. Then, for all $A\in \A$ we have
\be
\left\vert \omega(A)-\omega(PAP)\right\vert\leq 2\sqrt{1-\omega(P)}
\Vert A\Vert.
\ee
\end{lemma}
\begin{proof}
Note that
\begin{equation}
\omega(A)-\omega(PAP)=\omega(A(\idty-P))+\omega((\idty-P) A P).
\end{equation}
The two terms in the right-hand side can be estimated using 
the Cauchy-Schwarz inequality for the positive sesqui-linear form 
$(X,Y)\mapsto\omega(X^*Y)$: 
\bea
\vert \omega(A(\idty-P))\vert^2
&\leq& \omega(\idty-P)\omega(AA^*),\\
\vert \omega((\idty-P)AP)\vert^2
&\leq& \omega(\idty-P)\omega(PAA^*P).
\eea
As $\omega(AA^*)$ and $\omega(PAA^*P)$ are both bounded by $\Vert
A\Vert^2$, the result follows.
\end{proof}

The set of all states of $\A$ is w$*$-compact. Therefore, for
any state $\omega$ there are sequences $a_n$ of integers, with 
$\lim_n a_n=+\infty$, and such that the w$*$-limits of 
$\omega\circ\tau_{a_n}$ and $\omega\circ\tau_{-a_n}$ exist.
The following theorem is then sufficient to prove Theorem
\ref{thm:mainXXZ}.

\begin{theorem}
\label{thm:mainXXZbis}
Let $\omega$ be any ground state of the spin-S XXZ ferromagnetic chain 
with anisotropic coupling $\Delta>1$, and let $[a_n,b_n]$ be a sequence
of intervals tending to $\Ir$ (i.e., $a_n\to-\infty$ and $b_n\to+\infty$),
such that
\begin{equation}
\wslim_{n\rightarrow\infty}\omega\circ\tau_{-a_n}\equiv
\omega_{-\infty}=t_{-\infty}\omega_\uparrow+(1-t_{-\infty})\omega_\downarrow, 
\label{asymp1}
\end{equation}
and
\begin{equation}
\wslim_{n\rightarrow\infty}\omega\circ\tau_{b_n}\equiv
\omega_{+\infty}=t_{+\infty}\omega_\uparrow+(1-t_{+\infty})\omega_\downarrow.
\label{asymp2}\end{equation}
Here $t_{\pm\infty}\in[0,1]$. Then the following properties hold:
\par\noindent
i) $\omega$ has well-defined asymptotics, i.e.,
\begin{equation}
w^\ast\mbox{--}\lim_{n\rightarrow\infty}\omega\circ\tau_{\pm n}
=\omega_{\mp\infty}
\end{equation}
ii) $\omega$ has a convex decomposition as follows
\begin{equation}
\omega= t^{++}\omega_\uparrow + t^{--}\omega_\downarrow
+ t^{+-}\varphi^{+-} + t^{-+}\varphi^{-+}, 
\label{convexdecomp}\end{equation}
where the states $\varphi^{+-}$ and $\varphi^{-+}$ satisfy, for all
$x\in\Ir$,
\begin{equation}
\varphi^{+-}(h^{+-}_{x,x+1}) = 0,\quad \varphi^{-+}(h^{-+}_{x,x+1})=0. 
\end{equation}
\end{theorem}

Note that the complete list of such states $\varphi^{+-}$ and $\varphi^{-+}$
is known due to the work of Gottstein and Werner described in the previous 
section. The state $\varphi^{+-}$ is a convex combination of 
the translations of a kink state, and $\varphi^{-+}$ is a convex combination 
of the translations of an antikink state. The following relations between 
the convex combination coefficients are obvious: 
\begin{equation}
t_{-\infty} = t^{++} + t^{+-},\quad t_{+\infty} = t^{-+} + t^{++}.
\end{equation}

\begin{proof}
We use the Bratteli-Kishimoto-Robinson characterization of ground states
stated in Theorem \ref{thm:BKR}. A suitable choice for the boundary terms is:
\be
W_{[a,b]}=-2S A(\Delta)\left(P^+_{a-1}P^-_{b+1}+P^-_{a-1}P^+_{b+1}\right),
\label{defW}
\ee
where $P_x^{\pm}$ are the projections onto the state with $S_x^{(3)}=S$ at the 
site $x$ for the plus sign, and the state with $S_x^{(3)}=-S$ 
for the minus sign. The corresponding local Hamiltonian is 
\begin{equation}
{\tilde H}_{[a,b]}=\sum_{x=a-1}^b h_{x,x+1}+W_{[a,b]}. 
\end{equation}

First, we will show that for any ground state $\omega$
satisfying the assumptions of the theorem 
the quantity $E_{\Lambda}(\omega)$ defined by 
\be
E_{\Lambda}(\omega)=\inf\{\omega^\prime(\tilde H_\Lambda) \mid 
\omega^\prime_{\Lambda^c}=\omega_{\Lambda^c}\}
\label{E_vanishes}
\ee
tends to zero as $\Lambda=[a_n,b_n]\uparrow\Ir$ 
by showing that $\liminf_n E_{[a_n,b_n]}(\omega)
\geq 0$ (part a), and $\limsup_n E_{[a_n,b_n]}(\omega)\leq 0$ (part b).
This fact can then be combined with the assumed asymptotics (\ref{asymp1}
-\ref{asymp2}) to complete the proof of the theorem (see part c).
\medskip

{\it a) Proof of $\liminf_n E_{[a_n,b_n]}(\omega)\geq 0$.}\newline
We write 
\begin{equation}
P_x=P_x^++P_x^-.
\end{equation}
The assumed asymptotic behavior of $\omega$ implies that for any $\epsilon>0$, 
there exists $N$ such that, for all $n\geq N$, we have
\bea
\omega(\idty-P_{a_n-1})&\leq&\epsilon,\label{Pan1}\\
\omega(P^\alpha_{a_n-1}(\idty-P^\alpha_{a_n}))&\leq&\epsilon,\label{Pan2}\\
\omega(P^{\alpha}_{a_n-1}A_{a_n-1}P^{\beta}_{a_n-1})&\leq&\epsilon,
\quad\mbox{ if }\alpha\neq\beta \quad \mbox{for any} \ A_{a_n-1}\in
{\cal A}_{\{a_n-1\}}\label{Pan3}
\eea
and similar inequalities hold with $a_n-1$ replaced by $b_n+1$.
Clearly, for $\Lambda=[a_n,b_n],n\geq N$, any $\omega^\prime$ such
that $\omega^\prime_{\Lambda^c}=\omega_{\Lambda^c}$ satisfies the
same inequalities. In order to estimate the finite-volume energy of such
$\omega^\prime$, first introduce resolutions of the identity at the site 
$a_n-1$ as follows
\begin{eqnarray}
\omega^\prime(\tilde{H}_{[a_n,b_n]})
&=&\omega^\prime(P_{a_n-1}\tilde{H}_{[a_n,b_n]}P_{a_n-1})
+\omega^\prime(P_{a_n-1}\tilde{H}_{[a_n,b_n]}(\idty-P_{a_n-1}))\ret
&+&\omega^\prime((\idty-P_{a_n-1})\tilde{H}_{[a_n,b_n]}P_{a_n-1})
+\omega^\prime((\idty-P_{a_n-1})\tilde{H}_{[a_n,b_n]}(\idty-P_{a_n-1})).\ret
\label{decom1}
\end{eqnarray}
{From} (\ref{Pan1}), the non-vanishing terms are the first and 
the fourth ones. The fourth term is non-negative. In fact it is written as 
\begin{equation}
\omega'((\idty-P_{a_n-1})\tilde{H}_{[a_n,b_n]}(\idty-P_{a_n-1}))
=\sum_{x=a_n-1}^{b_n}\omega'((\idty-P_{a_n-1})h_{x,x+1}(\idty-P_{a_n-1}))\ge 0
\end{equation}
because the contribution from the boundary term $W_{[a_n,b_n]}$ of 
(\ref{defW}) is vanishing. We decompose the first term in the right-hand side 
of (\ref{decom1}) as 
\begin{eqnarray}
\omega^\prime(P_{a_n-1}\tilde{H}_{[a_n,b_n]}P_{a_n-1})
&=&\omega^\prime(P_{a_n-1}^+\tilde{H}_{[a_n,b_n]}P_{a_n-1}^+)
+\omega^\prime(P_{a_n-1}^+\tilde{H}_{[a_n,b_n]}P_{a_n-1}^-)\ret
&+&\omega^\prime(P_{a_n-1}^-\tilde{H}_{[a_n,b_n]}P_{a_n-1}^+)
+\omega^\prime(P_{a_n-1}^-\tilde{H}_{[a_n,b_n]}P_{a_n-1}^-). 
\end{eqnarray}
{From} (\ref{Pan3}), the only non-vanishing terms are of the form 
$\omega^\prime(P_{a_n-1}^\alpha\tilde{H}_{[a_n,b_n]}P_{a_n-1}^\alpha)$. 
Since similar properties hold for the site $b_n+1$, we have only to treat 
the forms 
\begin{equation}
\omega^\prime(P_{a_n-1}^\alpha P_{b_n+1}^\beta\tilde{H}_{[a_n,b_n]}
P_{a_n-1}^\alpha P_{b_n+1}^\beta)
\end{equation}
as the rest of non-vanishing contributions. 
We show that these are non-negative. To do this we consider the following 
decompositions of the Hamiltonian:
\be
{\tilde H}_{[a,b]}={\tilde H}_{[a,b]}^{\alpha\beta} 
+\delta W^{\alpha\beta}_{[a,b]}
\label{decompham}
\ee
with 
\begin{equation}
\delta W^{\alpha\beta}_{[a,b]}:=A(\Delta)
\left[\frac{1}{2}(\alpha-\beta)\left(S^{(3)}_{a-1}-S^{(3)}_{b+1}\right)
-2S\left(P^+_{a-1}P^-_{b+1}+P^-_{a-1}P^+_{b+1}\right)\right]. 
\label{deltaW}
\end{equation}
Using this expression, one can easily show 
\be
P^\alpha_{a-1}P^\beta_{b+1}\tilde H_{[a,b]}P^\alpha_{a-1}P^\beta_{b+1}
= P^\alpha_{a-1}P^\beta_{b+1}{\tilde H}^{\alpha\beta}_{[a,b]}
P^\alpha_{a-1}P^\beta_{b+1}\geq 0.
\ee
Clearly the corresponding terms in the expectation are non-negative. 
Thus, we have shown that 
\be
\omega(\tilde{H}_{[a_n,b_n]})\geq -C\times\epsilon, 
\ee
where $C$ is a positive constant and as $\epsilon>0$ is arbitrary, this
proves $\liminf_n E_{[a_n,b_n]}(\omega)\geq 0$.
\medskip

{\it b) Proof of $\limsup_n E_{[a_n,b_n]}(\omega)\leq 0$.}\newline
In order to obtain an upper bound implying that also $\limsup
E_{\Lambda}(\omega) =0$, we choose a trial state
$\omega^\prime$ defined as follows: 
\begin{equation}
\omega^\prime=\sum_{\alpha,\beta=\pm}\xi^{\alpha\beta}+\xi'
\end{equation}
with 
\begin{equation}
\xi^{\alpha\beta}=\psi^{\alpha\beta}_{[a_n+1,b_n-1]}
\otimes\omega_{[a_n+1,b_n-1]^c}(P^\alpha_{a_n}P^\beta_{b_n}(\ \cdots\ )
P^\alpha_{a_n}P^\beta_{b_n})
\end{equation}
and
\begin{equation}
\xi'=\psi^{++}_{[a_n+1,b_n-1]}\otimes\omega_{[a_n+1,b_n-1]^c}
((\idty-P_{a_n}P_{b_n})(\cdots)(\idty-P_{a_n}P_{b_n})),
\end{equation}
where $\psi^{-+}_{[a_n+1,b_n-1]}$ is given by 
(\ref{defkinkpsi}), i.e., the antikink centered at the origin, 
$\psi_{[a_n+1,b_n-1]}^{+-}$ is the kink which is the reflection of 
the kink, and $\psi^{\alpha\alpha}$ is the state 
with all spins $S_x^{(3)}=\alpha S$ for $\alpha=\pm$. 

Before making the energy estimates, we check that $\omega'(A)=\omega(A)$ 
for all $A\in{\cal A}_{[a_n,b_n]^c}$. This is verified as follows:
\begin{eqnarray}
\omega'(A)&=&\sum_{\alpha,\beta=\pm}\xi^{\alpha\beta}(A)+\xi'(A)\ret
&=&\sum_{\alpha,\beta=\pm}\omega(P_{a_n}^\alpha P_{b_n}^\beta A)
+\omega((\idty-P_{a_n}P_{b_n})A)\ret
&=&\omega(P_{a_n}P_{b_n}A)+\omega((\idty-P_{a_n}P_{b_n})A)=\omega(A)
\end{eqnarray}
for $A\in{\cal A}_{[a_n,b_n]^c}$.

To compute the energy of each term we again use the decomposition 
(\ref{decompham}). The interior terms of the Hamiltonian, i.e., 
${\tilde H}^{\alpha\beta}_{[a_n+2,b_n-2]}$, are identically zero. 
Therefore we have 
\begin{eqnarray}
\xi^{\alpha\beta}\left({\tilde H}_{[a_n,b_n]}\right)
&=&\xi^{\alpha\beta}\left(h_{a_n-1,a_n}^{\alpha\beta}
+h_{a_n,a_n+1}^{\alpha\beta}
+h_{b_n-1,b_n}^{\alpha\beta}+h_{b_n,b_n+1}^{\alpha\beta}\right)\ret
&+&\xi^{\alpha\beta}\left(\delta W_{[a_n,b_n]}^{\alpha\beta}\right),
\label{xialphabetaest1}
\end{eqnarray}
and 
\begin{eqnarray}
\xi'\left({\tilde H}_{[a_n,b_n]}\right)&=&
\xi'\left(h_{a_n-1,a_n}+h_{a_n,a_n+1}+h_{b_n-1,b_n}+h_{b_n,b_n+1}\right)\ret
&-&2SA(\Delta)\xi'\left(P_{a_n-1}^+P_{b_n+1}^-+P_{a_n-1}^-P_{b_n+1}^+\right).
\label{xipest}
\end{eqnarray}
Owing to (\ref{Pan1}) and 
\begin{equation}
\idty-P_{a_n}P_{b_n}=(\idty-P_{a_n})P_{b_n}+P_{a_n}(\idty-P_{b_n})
+(\idty-P_{a_n})(\idty-P_{b_n}), 
\end{equation}
the right-hand side of (\ref{xipest}) tends to zero as $n\to\infty$. 

By definition, we can write 
$\xi^{\alpha\beta}\left({\tilde H}_{[a_n,b_n]}\right)$ 
of (\ref{xialphabetaest1}) as 
\begin{eqnarray}
\xi^{\alpha\beta}\left({\tilde H}_{[a_n,b_n]}\right)
&=&\omega\left(P_{b_n}^\beta P_{a_n}^\alpha h_{a_n-1,a_n}^{\alpha\beta}
P_{a_n}^\alpha\right)
+\xi^{\alpha\beta}(h_{a_n,a_n+1}^{\alpha\beta})
+\xi^{\alpha\beta}(h_{b_n-1,b_n}^{\alpha\beta})\ret&+&
\omega\left(P_{a_n}^\alpha P_{b_n}^\beta h_{b_n-1,b_n}^{\alpha\beta}
P_{b_n}^\beta\right)
+\omega\left(P_{a_n}^\alpha P_{b_n}^\beta\delta 
W_{[a_n,b_n]}^{\alpha\beta}\right).
\label{xialphabetaest}
\end{eqnarray}
The first term in the right-hand side is written as 
\begin{eqnarray}
& &\omega\left(P_{b_n}^\beta P_{a_n}^\alpha h_{a_n-1,a_n}^{\alpha\beta}
P_{a_n}^\alpha\right)\ret
&=&\omega\left(P_{b_n}^\beta P_{a_n}^\alpha h_{a_n-1,a_n}^{\alpha\beta}
P_{a_n}^\alpha(\idty-P_{a_n-1}^\alpha)\right)
+\omega\left(P_{b_n}^\beta P_{a_n}^\alpha h_{a_n-1,a_n}^{\alpha\beta}
P_{a_n}^\alpha P_{a_n-1}^\alpha\right).
\end{eqnarray}
This first term in the right-hand side is vanishing as 
$a_n\rightarrow -\infty$ from (\ref{Pan2}), and the second term also 
is vanishing because $h_{a_n-1,a_n}^{\alpha\beta}P_{a_n}^\alpha 
P_{a_n-1}^\alpha=0$. Therefore the first term 
$\omega\left(P_{b_n}^\beta P_{a_n}^\alpha h_{a_n-1,a_n}^{\alpha\beta}
P_{a_n}^\alpha\right)$ 
of (\ref{xialphabetaest}) is vanishing. 
Similarly the fourth term in the right-hand side of (\ref{xialphabetaest}) is 
vanishing as $b_n\rightarrow +\infty$. 

The fifth term in the right-hand side of (\ref{xialphabetaest}) tends to zero 
as $n\to\infty$. To show this, we introduce a resolution of the identity 
\begin{equation}
\idty=P_{a_n-1}^\alpha P_{b_n+1}^\beta+P_{a_n-1}^\alpha(\idty-P_{b_n+1}^\beta)
+(\idty-P_{a_n-1}^\alpha)P_{b_n+1}^\beta+(\idty-P_{a_n-1}^\alpha)
(\idty-P_{b_n+1}^\beta). 
\end{equation}
Owing to this and (\ref{Pan2}), we have only to show that 
\begin{equation}
\omega\left(P_{a_n}^\alpha P_{b_n}^\beta
\delta W_{[a_n,b_n]}^{\alpha\beta} P_{a_n-1}^\alpha P_{b_n+1}^\beta\right)
\end{equation}
is vanishing in the limit $n\to\infty$. But this is identically zero for all 
$\alpha,\beta=\pm$. 

The rest are the second and the third terms in the right-hand side of 
(\ref{xialphabetaest}). We treat only the second term 
because the third one is treated in the same way. To begin with, 
we note the following: {From} Lemma~\ref{lem:probnotS} and 
Lemma~\ref{lem:state_bound} we obtain 
\begin{equation}
\left\Vert\psi^{\alpha\beta}_{[a_n+1,b_n-1]}(P^\alpha_{a_n+1} 
P^\beta_{b_n-1}(\ \cdots\ )
P^\alpha_{a_n+1}P^\beta_{b_n-1})-\psi^{\alpha\beta}_{[a_n+1,b_n-1]}(\cdots)
\right\Vert
\leq\mu\exp\left[-\nu\min\{-a_n,b_n\}\right]
\label{psiasym}
\end{equation}
for some positive constants $\mu, \nu$, and $\min\{-a_n,b_n\}$ represents 
the distance of the position of the center of the kink to the origin. 
Combining (\ref{psiasym}) with 
\begin{equation}
P^\alpha_{a_n}P^\alpha_{a_n+1}h^{\alpha\beta}_{a_n,a_n+1}P^\alpha_{a_n} 
P^\alpha_{a_n+1}=0
\quad \mbox{for all} \ \alpha,\beta=\pm, 
\end{equation}
we get the desired result 
$\xi^{\alpha\beta}(h_{a_n,a_n+1}^{\alpha\beta})\rightarrow 0$ as 
$a_n\rightarrow -\infty$. 
This concludes the proof of $\limsup_n E_{[a_n,b_n]}(\omega)\leq 0$.
\medskip

{\it c)} We can now complete the proof as follows.
{From} the asymptotics it follows that there is a sequence $\epsilon_n
\downarrow 0$ such that
\begin{equation}
\bigl\vert \omega_{[a_n,b_n]}(A)-\sum_{\alpha,\beta=\pm}
\omega(P^\alpha_{a_n -1}P^\beta_{b_n+1} A)\bigr\vert\leq \epsilon_n
\Vert A\Vert
\end{equation}
for all $A\in\A_{[a_n,b_n]}$. This implies that there are four subsequences
of states defined by
\begin{equation}
\omega^{\alpha\beta}_k(A)
=\frac{\omega(P^\alpha_{a_{n_k} -1}P^\beta_{b_{n_k}+1}A P^\alpha_{a_{n_k} 
-1}P^\beta_{b_{n_k}+1})}{\omega(P^\alpha_{a_{n_k} -1}P^\beta_{b_{n_k}+1})}
\end{equation}
with the following properties:
\begin{eqnarray}
&&w^\ast\mbox{--}\lim_{k\rightarrow\infty}\omega^{\alpha\beta}_k 
\equiv\omega^{\alpha\beta}\ \mbox{exists}\label{property1},\\
&&w^\ast\mbox{--}\lim_{k\rightarrow\infty}\left\vert\omega
-\sum_{\alpha,\beta=\pm} t^{\alpha\beta} \omega^{\alpha\beta}_k\right\vert
=0, \quad \mbox{with} \ t^{\alpha\beta}\ge 0, \label{property3}\\
&&\lim_{k\rightarrow\infty}\omega^{\alpha\beta}_k(h_{x,x+1}^{\alpha\beta})=0
\quad \mbox{for all $x\in \Ir$}.\label{property2} 
\end{eqnarray}
The last property (\ref{property2}) follows from the above results 
in parts a and b. To see this, we note that 
\begin{eqnarray}
\omega_k^{\alpha\beta}\left({\tilde H}_{[a_{n_k},b_{n_k}]}\right)
&=&\omega_k^{\alpha\beta}\left({\tilde H}_{[a_{n_k},b_{n_k}]}^{\alpha\beta}
\right)
+\omega_k^{\alpha\beta}\left(\delta W_{[a_{n_k},b_{n_k}]}^{\alpha\beta}\right),
\end{eqnarray}
where we have used the decompositions (\ref{decompham}) 
of the Hamiltonian ${\tilde H}_{[a,b]}$. The second term in 
the right-hand side is vanishing as $a_{n_k}\rightarrow -\infty,
b_{n_k}\rightarrow +\infty$ from the asymptotics and the definition 
(\ref{deltaW}) of $\delta W_{[a,b]}^{\alpha\beta}$. 
Combining these observations with 
(\ref{property3}) and with the results of parts a and b, i.e., 
$\omega({\tilde H}_{[a_n,b_n]})\rightarrow 0$ as $a_n\rightarrow -\infty, 
b_n\rightarrow +\infty$, we have 
\begin{equation}
\omega_k^{\alpha\beta}({\tilde H}_{[a_{n_k},b_{n_k}]}^{\alpha\beta})
=\sum_{x=a_{n_k}-1}^{b_{n_k}} \omega_k^{\alpha\beta}(h_{x,x+1}^{\alpha\beta})
\rightarrow 0 \quad \mbox{as} \ a_{n_k}\rightarrow -\infty, 
b_{n_k}\rightarrow +\infty.
\end{equation}
This implies (\ref{property2}) because $h_{x,x+1}^{\alpha\beta}\ge 0$ 
for all $\alpha,\beta=\pm$ and for all $x\in \Ir$.

As the convex decomposition (\ref{convexdecomp}), if it exists, is 
unique, its terms must be proportional to the $\omega^{\alpha\beta}$
because of (\ref{property3}).
The property (\ref{property2}) identifies the $\omega^{\alpha\beta}$
as the known zero-energy states, which proves the statement ii) of 
Theorem~\ref{thm:mainXXZbis}. The statement i) follows from ii).
\end{proof}

\section[PTXXX]{Proof of Theorem \ref{thm:mainXXX}}
\label{sec:proofXXX}
\setcounter{equation}{0}%
\setcounter{theorem}{0}%

The proof of the absence of non-translation-invariant ground states
in the isotropic case is similar to the proof in the anisotropic case
(Theorem \ref{thm:mainXXZbis}). It differs from it in two points.
The first difference is that in the isotropic case all ground states
turn out to be zero energy ground states for the {\em same} local 
Hamiltonians. This simplifies the proof. The second difference, however,
makes the proof more subtle than in the anisotropic case. This is due to
the broken continuous rotation symmetry. The possible
asymptotics of pure ground states now depend on two continuous parameters,
which can be taken to be two angles: $(\theta,\phi)=\Omega\in S^2$.
For the same reason the excitation spectrum is gapless and, therefore,
the method of proof followed in \cite{Mat} cannot be adapted to the 
isotropic case.

The state of a single spin pointing in the direction $\Omega$ is represented 
by the vector 
\begin{equation}
\ket{\Omega}:=U(\Omega)\ket{S}, 
\end{equation}
where
\be
U(\Omega):=e^{-i\phi S^{(3)}}e^{-i\theta S^{(2)}}, 
\ee
and $\ket{S}$ is the normalized eigenvector of $S^{(3)}$ satisfying 
$S^{(3)}\ket{S}=S\ket{S}$. It follows that the vectors
\begin{equation}
U(\Omega)\ket{S},\quad \Omega\in S^2
\end{equation} 
span the $(2S+1)$-dimensional irreducible unitary representation of SU(2).
For the same reason the vectors
\begin{equation}
\ket{\Omega}_\Lambda:=\bigotimes_{x\in\Lambda}(\ket{\Omega})_x
\end{equation}
span the maximum total spin subspace for any finite volume $\Lambda\subset\Ir$.
It is also straightforward to check that the orthogonal projection
onto the subspace of maximal total spin, $P_\Lambda$, can be written as 
\cite{KlaSka}
\be
P_\Lambda:=\frac{2kS+1}{4\pi}\int d\Omega\,Q_\Lambda(\Omega), 
\label{defP}
\ee
where, as usual, $d\Omega=\sin \theta d\theta d\phi$, and
\be
Q_\Lambda(\Omega)=\bigotimes_{x\in\Lambda}(\ket{\Omega}\bra{\Omega})_x. 
\label{defQ}\ee

It is obvious from the form of the isotropic Hamiltonian (\eq{Ham_ppmm}  with
$\Delta =1$) that  zero-energy states are supported by the maximum total spin
subspace on each finite volume, which is permutation invariant. It follows
immediately that all zero-energy ground states of this Hamiltonian are
translation invariant. Therefore, for the proof of Theorem \ref{thm:mainXXX} it is
sufficient to show that all ground states are zero-energy states for this
Hamiltonian. By Theorem \ref{thm:BKR}, and the fact that the local
interactions are non-negative, this will follow if we prove that for any
ground state
\be
\lim_{\Lambda\uparrow\Ir}E_\Lambda(\omega)=0, 
\label{energy_XXX_vanishes}\ee
where $E_\Lambda(\omega)$ is defined in \eq{E_vanishes}. 
By Lemma~\ref{lem:finite_energy} we have to do this for any $\omega$ 
with left and right asymptotics such that
\be
\lim_{x\to\pm\infty}\omega(\idty-P_{\Lambda +x})=0
\label{asymXXX}
\ee
for all finite $\Lambda\subset\Ir$. To do this we follow the same strategy
as in the proof of Theorem \ref{thm:mainXXZ}: we need to construct local
modifications of an arbitrary ground state $\omega$ that have 
arbitrarily low energy. This will be done by inserting a long-wavelength
spin-wave state that gradually turns the spin from $\Omega^-$
to $\Omega^+$, conditioned upon the asymptotic orientation to the left 
and to the right being $\Omega^-$ and $\Omega^+$ respectively. The sum over 
the possible asymptotic behaviors appearing in the proof of Theorem
\ref{thm:mainXXZ} will here become an integral over $\Omega^-$ and $\Omega^+$.
Some care has to be taken with  the conditioning in order not to spoil the
energy estimates, which need to be done carefully, too. A quick estimate
produces useless bounds. Now, we fill in the technical details of the proof 
sketched above.

\vspace{.5truecm}

\noindent
{\bf Proof of Theorem \ref{thm:mainXXX}:}\newline
\noindent
As explained above we want to show \eq{energy_XXX_vanishes}.
Due to the non-negative interaction we only need to show
\be
\limsup_{\Lambda\uparrow\Ir}E_\Lambda(\omega)\leq 0. 
\label{limsup_XXX}\ee
Let $\omega$ be a ground state. We will construct
a trial state $\omega^\prime$ which coincides with $\omega$
outside a finite interval $\Lambda=[a,b]$. We will use a corridor
of $m$ sites to perform the conditioning. Therefore, let $m\geq 1$ be
such that $b-a+1>2m$. The state $\omega^\prime$ is then defined by
\begin{equation}
\omega':=\int d\Omega^{+}\int d\Omega^{-}
\eta_{[a+m,b-m]}^{\Omega^{+},\Omega^{-}}\otimes 
\omega_{{[a+m,b-m]}^c}^{\Omega^{+},\Omega^{-}}+\omega^\pprime,
\label{defomegap}
\end{equation}
where $\eta_{[a+m,b-m]}^{\Omega^{+},\Omega^{-}}$,
$\omega_{{[a+m,b-m]}^c}^{\Omega^{+},\Omega^{-}}$, and $\omega^\pprime$
are non-negative functionals on $\A_{[a+m,b-m]}$, \hfill\break 
$\A_{{[a+m,b-m]}^c}$, and $\A$, respectively, defined as follows. 
For convenience put ${\hat \Lambda}=[a+m,b-m]$.
\begin{equation}
\eta_{\hat \Lambda}^{\Omega^{+},\Omega^{-}}(\,\cdots\,):=
\omega_{\uparrow,{\hat \Lambda}}\left(\left(
V_{\hat \Lambda}^{\Omega^{+},\Omega^{-}}\right)^*
(\,\cdots\,)V_{\hat \Lambda}^{\Omega^{+},\Omega^{-}}\right),
\label{defvarphi}
\end{equation}
where $\omega_{\uparrow,\hat{\Lambda}}$ is the state with 
$S^{(3)}_x=S$ for all $x\in\hat\Lambda$,
and $V_{\hat \Lambda}^{\Omega^+,\Omega^-}$ is the unitary defined by
\begin{eqnarray}
V_{\hat \Lambda}^{\Omega^+,\Omega^-}&:=&
\exp\left[-i\phi^-\sum_{x=a+m}^{b-m}S_x^{(3)}-\frac{i}{\hat L}
\left(\phi^+-\phi^-\right)
\sum_{x=a+m}^{b-m}(x-a-m)S_x^{(3)}\right] \nonumber \\
&\times& \exp\left[-i\theta^-\sum_{x=a+m}^{b-m}S_x^{(2)}-\frac{i}{\hat L}
\left(\theta^+-\theta^-\right)\sum_{x=a+m}^{b-m}(x-a-m)
S_x^{(2)}\right] 
\end{eqnarray}
with ${\hat L}=b-a-2m$, and $\Omega^\pm=(\theta^\pm,\phi^\pm)$. 
\bea
&&\omega_{{\hat \Lambda}^c}^{\Omega^{+},\Omega^{-}}(\cdots)
:=\left(\frac{2mS+1}{4\pi}\right)^2\nonumber\\
&&\times\omega_{{\hat \Lambda}^c}
\left(Q_{[a,a+m-1]}(\Omega^-) Q_{[b-m+1,b]}(\Omega^+)(\,\cdots\,)
Q_{[a,a+m-1]}(\Omega^-) Q_{[b-m+1,b]}(\Omega^+)\right),
\label{omegapro}\eea
where $Q_\Lambda(\Omega)$ is defined in \eq{defQ}.
\begin{equation}
\omega''(\,\cdots\,):=\omega_{\uparrow, \hat \Lambda}(\,\cdots\,)\otimes 
\omega_{{\hat \Lambda}^c}\left((\idty-P_{[a,a+m-1]}P_{[b-m+1,b]})
(\,\cdots\,)(\idty-P_{[a,a+m-1]}P_{[b-m+1,b]})\right),
\label{defomegapp}
\end{equation}
with $P_{[k,\ell]}$ defined in \eq{defP}.

Note that in the state $\eta_{\hat \Lambda}^{\Omega^{+},\Omega^{-}}$ 
the orientation of spins gradually rotate from $\Omega^{-}$ 
on the left to $\Omega^{+}$ on the right. 
Due to the known asymptotics of any ground state,
the state $\omega_{{\hat \Lambda}^c}^{\Omega^{+},\Omega^{-}}$,
up to normalization, represents $\omega$ conditioned upon
the spins having orientation $\Omega^-$ on the sites $a,\ldots, a+m-1$
and  $\Omega^+$ on the sites $b-m+1,\ldots,b$.
In finite volume all this is approximate, and the term 
$\omega''$ is exactly the correction needed to reproduce $\omega$ 
on the complement of $\Lambda$. 

Before making the energy estimates, we first verify this fact, 
i.e., check that $\omega'(A)=\omega(A)$ for all $A\in {\cal A}_{\Lambda^c}$. 
Note that the projection operators $P_{[a,a+m-1]}$, $P_{[b-m+1,b]}$,
\hfill\break 
$Q_{[a,a+m-1]}(\Omega^-)$, and $Q_{[b-m+1,b]}(\Omega^+)$ are elements of 
the algebra ${\cal A}_{\Lambda \cap {\hat \Lambda}^c}$, 
and therefore commute with any $A\in \A_{\Lambda^c}$ by the definitions. 
This fact and the definitions given above suffice 
to check the following for all $A\in\A_{\Lambda^c}$:
\begin{eqnarray}
\omega'(A)&=&\int d\Omega^+\int d\Omega^-\,
\eta_{\hat \Lambda}^{\Omega^+,\Omega^-}(\idty)
\omega_{{\hat \Lambda}^c}^{\Omega^{+},\Omega^{-}}(A)
+\omega''(A)\ret
&=&\left(\frac{2mS+1}{4\pi}\right)^2\int d\Omega^+\int d\Omega^-\,
\omega(Q_{[a,a+m-1]}(\Omega^-) Q_{[b-m+1,b]}(\Omega^+)A)\ret
&+&\omega((\idty-P_{[a,a+m-1]}P_{[b-m+1,b]})A)\ret
&=&\omega(P_{[a,a+m-1]}P_{[b-m+1,b]}A)
+\omega((\idty-P_{[a,a+m-1]}P_{[b-m+1,b]})A)=\omega(A).
\end{eqnarray}

Next, we estimate the energy of $\omega^\prime$. By definition, we have 
\begin{eqnarray}
\omega'(H_{\Lambda\cup\partial\Lambda})&=&\int d\Omega^+\int d\Omega^-
\eta_{\hat \Lambda}^{\Omega^+,\Omega^-}
\left(H_{\hat \Lambda}\right)\nonumber \\
&+&\int d\Omega^+\int d\Omega^-
\left[\omega_{{\hat \Lambda}^c}^{\Omega^+,\Omega^-}(h_{a-1,a})
+\omega_{{\hat \Lambda}^c}^{\Omega^+,\Omega^-}(h_{b,b+1})\right]
\nonumber \\
&+&\sum_{x=b-m}^b\omega''(h_{x,x+1})+\sum_{x=a-1}^{a+m-1}
\omega''(h_{x,x+1}).
\label{omegapenergy}
\end{eqnarray}

First, we estimate $\eta_{\hat \Lambda}^{\Omega^+,\Omega^-}
\left(H_{\hat\Lambda}\right)$. Due to the rotation invariance 
this quantity depends only on the angle between $\Omega^-$ and 
$\Omega^+$. Therefore we can consider the case
$\theta^+=\theta,\theta^-=\phi^+=\phi^-=0$. The energy is then given
by the usual spin wave energy: 
\begin{eqnarray}
\eta_{\hat \Lambda}^{\Omega^+,\Omega^-}\left(H_{\hat \Lambda}\right)
&=&\sum_{x=a+m}^{b-m-1}\omega_\uparrow
\left(\left(V_{\hat \Lambda}^{\Omega^+,\Omega^-}\right)^\ast
h_{x,x+1}V_{\hat \Lambda}^{\Omega^+,\Omega^-}\right)
\nonumber \\
&=&S^2{\hat L}\left[1-\cos(\theta/{\hat L})\right] . 
\end{eqnarray}
This implies 
\be
\eta_{\hat \Lambda}^{\Omega^+,\Omega^-}\left(H_{\hat \Lambda}\right)
\rightarrow 0, \mbox{ as }{\hat L}=b-a-2m \rightarrow \infty .
\label{est1}\ee
Therefore the first integral 
in the right-hand side of (\ref{omegapenergy}) vanishes
in the limit ${\hat \Lambda}\uparrow\Ir$. 

Next consider the second integral in the right-hand side of 
(\ref{omegapenergy}). Using the definitions (\ref{defP}), (\ref{defQ}), 
and (\ref{omegapro}), we have 
\begin{eqnarray}
& &\int d\Omega^+\int d\Omega^-
\omega_{{\hat \Lambda}^c}^{\Omega^+,\Omega^-}(h_{b,b+1})\nonumber\\
&=&\frac{2mS+1}{4\pi}\int d\Omega^+ \omega(Q_{[b-m+1,b]}(\Omega^+)
h_{b,b+1}Q_{[b-m+1,b]}(\Omega^+)P_{[a,a+m-1]})
\nonumber\\
&\le&\omega\left(\frac{2mS+1}{4\pi}\int d\Omega^+ Q_{[b-m+1,b]}(\Omega^+)
h_{b,b+1}Q_{[b-m+1,b]}(\Omega^+)\right).
\label{term2}\end{eqnarray}
The operator
\begin{equation}
\frac{2mS+1}{4\pi}\int d\Omega^+ Q_{[b-m+1,b]}(\Omega^+)
h_{b,b+1}Q_{[b-m+1,b]}(\Omega^+)
\end{equation}
commutes with SU(2) rotations. Therefore, as any weak limit 
$\omega \circ\tau_b$, $b\to\infty$, is supported by the highest
spin irreducible representation of SU(2), the limit $b\to\infty$
of \eq{term2} is given by
\begin{eqnarray}
& &\frac{2mS+1}{4\pi}\int d\Omega \,\omega^{(+\infty)}
\left(Q_{[1,m]}(\Omega)h_{m,m+1}Q_{[1,m]}(\Omega)\right)\nonumber\\
&=&\frac{2mS+1}{4\pi}\int d\Omega \,\omega_\uparrow
\left(Q_{[1,m]}(\Omega)h_{m,m+1}Q_{[1,m]}(\Omega)\right).
\label{est2}
\end{eqnarray}
Here again $\omega_\uparrow$ is the state determined by 
$\omega_\uparrow(S_x^{(3)})=S$, for all $x\in \Ir$. The right-hand side of 
\eq{est2} can easily be calculated: 
\begin{eqnarray}
& &\frac{2mS+1}{4\pi}\int d\Omega \,\omega_\uparrow
\left(Q_{[1,m]}(\Omega)h_{m,m+1}Q_{[1,m]}(\Omega)\right)\nonumber\\
&=&\frac{2mS+1}{4\pi}\int d\Omega \ S^2 
\left(\cos \frac{\theta}{2}\right)^{4mS}(1-\cos \theta)\nonumber\\
&=&\frac{S^2(2mS+1)}{2}\int_0^\pi d\theta \sin \theta 
\left(\frac{1+\cos\theta}{2}\right)^{2mS}(1-\cos \theta)=
\frac{S^2}{mS+1}. 
\label{cosint}
\end{eqnarray}
Clearly this quantity vanishes as $m\to \infty$. The contribution
in the second term of \eq{omegapenergy} coming form the 
left asymptotics can be estimated in the same way.
Therefore, we have shown 
\begin{equation}
\lim_{m \rightarrow \infty}
\lim_{b \rightarrow \infty} \lim_{a \rightarrow -\infty}
\int d\Omega^+\int d\Omega^-
\left[\omega_{{\hat \Lambda}^c}^{\Omega^+,\Omega^-}(h_{a-1,a})
+\omega_{{\hat \Lambda}^c}^{\Omega^+,\Omega^-}(h_{b,b+1})\right]=0 .
\label{cosint2}
\end{equation}

Finally, we consider the two summations in the right-hand side of 
\eq{omegapenergy}. Note that 
\begin{eqnarray}
\idty-P_{[a,a+m-1]}P_{[b-m+1,b]}
&=&P_{[a,a+m-1]}(\idty-P_{[b-m+1,b]})+(\idty-P_{[a,a+m-1]})P_{[b-m+1,b]}\ret
&+&(\idty-P_{[a,a+m-1]})(\idty-P_{[b-m+1,b]}).
\end{eqnarray}
Combining this with \eq{asymXXX} and \eq{defomegapp}, we can conclude 
\begin{equation}
\lim_{b\rightarrow \infty}\lim_{a\rightarrow -\infty}
\sum_{x=b-m}^b\omega''(h_{x,x+1})=0 ,\mbox{ and }
\lim_{b\rightarrow \infty}\lim_{a\rightarrow -\infty}
\sum_{x=a-1}^{a+m-1}\omega''(h_{x,x+1})=0 
\label{est3}\end{equation}
for a fixed $m$. 

Combining \eq{omegapenergy}, \eq{est1}, \eq{cosint2}, and \eq{est3}, we obtain 
\begin{equation}
\lim_{m \rightarrow \infty}
\lim_{b \rightarrow \infty} \lim_{a \rightarrow -\infty}
\omega'(H_{\Lambda\cup\partial\Lambda})=0 . 
\end{equation}
This implies \eq{limsup_XXX}.
\QED

\bigskip
\bigskip

\noindent
{\Large\bf Acknowledgements} 
\medskip

\noindent
It is a pleasure to thank the following people for discussions 
and correspondence:
Mieke De Cock, Mark Fannes, Taku Matsui, Seiji Miyashita, 
Jan-Philip Solovej, Hal Tasaki, and Reinhard Werner. 
B.N. was partially supported by the National Science Foundation 
under grant \# DMS-9706599.

\end{document}